\documentstyle[epsfig]{aipproc}

\def\ltsima{$\; \buildrel < \over \sim \;$}
\def\gtsima{$\; \buildrel > \over \sim \;$}
\def\lsim{\lower.5ex\hbox{\ltsima}}
\def\gsim{\lower.5ex\hbox{\gtsima}}
\begin{document}
\title{Searching for Millisecond Pulsars in Globular Clusters at Parkes:
Further Results}

\author{\vskip -0.4truecm
Andrea Possenti$^1$, Nichi D'Amico$^1$, Richard N. Manchester$^2$, 
John Sarkissian$^3$, Andrew G. Lyne$^4$ and Fernando Camilo$^5$}
\address{\vskip -0.4truecm
$^1$Osservatorio Astronomico di Bologna, via Ranzani 1, 
40127 Bologna, Italy}
\address{\vskip -0.7truecm
$^2$Australia Telescope National Facility,
CSIRO, PO Box 76, Epping, NSW 2121, Australia}
\address{\vskip -0.7truecm
$^3$Australia Telescope National Facility,
CSIRO, Parkes Observatory, PO Box 276, \\
Parkes, NSW 2870, Australia}
\address{\vskip -0.7truecm
$^4$University of Manchester, Jodrell Bank
Observatory, Macclesfield, SK11~9DL, UK}
\address{\vskip -0.7truecm
$^5$Columbia Astrophysics
Laboratory, Columbia University, 550 West 120th Street, \\
New York, NY 10027}

\maketitle
\vskip -0.9truecm
\begin{abstract}
We have discovered 12 new millisecond pulsars in 6 globular clusters in which 
no pulsars were previously known, in
the first two years of a search at 1.4 GHz in progress at the Parkes
radio telescope. Here we briefly describe the motivation, the
new hardware and software systems adopted for this survey,
and we present the results obtained thus far.
\end{abstract}

\section*{}
\begin{center}
\vskip -1.0truecm
{\sl Talk presented at the ACP Summer 2001 Workshop \\ 
``Compact Objects in Dense Star Clusters''\\ Aspen, 26 June 2001}
\end{center}

\section*{Introduction}

Millisecond pulsars (MSPs) are formed in binary systems containing 
a neutron star (NS) which is eventually spun up through mass
accretion from the evolving companion \cite{sb76,bv91,ka96}.
Despite the large difference in total mass between the disk of the Galaxy
and the Globular Cluster (GC) system, about 50\% of the entire 
MSP population has been found in the latter.
This is not surprising because, apart from the evolution of primordial
binaries, another formation channel for the MSPs is available in these 
stellar systems: exchange interactions in the ultra-dense stellar 
environment of the cluster core can sustain the formation of various kinds
of binaries suitable for recycling neutron stars \cite{dh98,rprp01}. 
  
Many authors (e.g. \cite{phi92,hmg+92,bv91b,fcl+00})
have discussed the importance of the discovery of GC-MSPs as diagnostic
tools for studying the dynamics of the clusters, the evolution of the binaries, 
the interstellar medium and the intracluster ionized gas
\cite{fklcmd01} but, except for the significant case of 47~Tuc \cite{clf+00},
the discovery rate of these objects strongly declined in the second
half of the 1990s: in the 7 years from 1987 (when B1821$-$24
was discovered in M28 at Jodrell Bank \cite{lbmkb87}) to 1994 \cite{bbfglb94}
32 GC-MSPs entered the catalog, whereas no new source was
published in the following 5 years.  
 
Passing reasons (e.g. the upgrade of the Arecibo telescope) may have
contributed to this trend, but a more meaningful explanation is that
{\it searches for GC-MSPs are difficult because they are often distant
pulsars in close binary systems\/}. Their large distances {\it (i)} make their
fluxes typically very small and {\it (ii)} strongly distort their signals due 
to the dispersive effects of propagation through the interstellar medium.
Their inclusion in tight binaries {\it (iii)} causes Doppler-shift changes
of the apparent spin period and sometimes {\it (iv)}
makes the radio signal periodically obscured by eclipses.  

In the last couple of years a new search still in progress at the
Parkes radio telescope has broken the hiatus in discoveries.
In the following we review the current status of this
experiment, including the preliminary announcement
of two new clusters with which an MSP has now been associated \cite{pdm+01}.
\vspace{-0.1truecm}

\section*{\vspace{-0.05truecm}Observations and Analysis}

The availability of a new highly sensitive 20-cm receiver at Parkes 
together with a modern data acquisition system
impelled us to undertake a new search of GCs for MSPs.  
This receiver has a system temperature of $\sim 21$~K and a bandwidth 
of $\sim$ 300 MHz: performing long integrations ($\sim 1-2$ hr) 
we have been able to reach a very good nominal sensitivity (at signal-to-noise
ratio s/n=8)
of $\sim 100-150~\mu$Jy for a typical 3 ms pulsar with 
dispersion measure DM $\sim 100-200$ cm$^{-3}$ pc. 

With the aim of improving our capability for probing distant clusters, 
we have designed and assembled at Jodrell Bank and Bologna
a high resolution filterbank system consisting of 
$512\times 0.5$ MHz adjacent channels per polarization.  This enables us
to minimize the deleterious effects of dispersion in the interstellar medium
sufficiently to maintain significant sensitivity to MSPs with 
DM $\lsim 300$ cm$^{-3}$pc. 

We have selected about 60 clusters among the ones visible at Parkes
according to their optical central density and satisfying 
the requirement DM$_{\rm exp} \le 300$ cm$^{-3}$pc (where DM$_{\rm exp}$
is the DM expected for the cluster according to a model
for the Galactic distribution of the ionized gas \cite{tc93}):
only a few of them contain already known MSPs.
In order to reduce the probability of missing strong signals temporarily
obscured by eclipse events we have performed multiple
observations of each target cluster.

After adding the outputs of 1024 channels in polarization pairs,
the resulting 512 data streams are each integrated and 1-bit digitized
every 125 $\mu$s: thus each observation produces a huge 
array, typically of $2-8$ Gbytes, requiring significant CPU 
resources for offline processing. In Bologna, we use a local cluster of 
10 Alpha-500MHz CPUs and on occasion the Cray-T3E 256-processor system at 
the CINECA Supercomputing Center. 

In the offline processing, each data stream
is split into non-overlapping segments of 2100, 4200 or 8400 sec and these
are processed separately.  When no pulsar is known in a GC (and so
the DM is unknown) the data are first de-dispersed over a wide range 
of $\sim 500-1000$ trial DMs, spanning the interval DM$_{\rm exp}\pm$40\%.
Each de-dispersed series is then transformed using a Fast Fourier 
Transform. In the subsequent step, 
time-domain data are folded in sub-integrations using constant
periods, each corresponding to a large number of spectral features above 
a given threshold. The resulting ``sub-integration arrays'' are 
searched both for a linear and for a parabolic shift in pulse phase.
The linear correction is a consequence of having folded the data
at an approximate spin period, whereas the parabolic correction
is a signature of the acceleration of the source due to its orbital
motion. Parameters for final pulse profiles with significant s/n 
are displayed for visual inspection. 

This processing scheme gives some sensitivity
for MSPs belonging to binary systems and most of our discoveries
have been obtained using it. However, once a pulsar 
is detected and confirmed in a cluster, we usually reprocess
the data de-dispersing them at the single DM value of the newly 
discovered pulsar; then the resulting time series 
undergoes a fully coherent search for ``accelerated'' signals
over a large range of acceleration values. Being extremely CPU
time-consuming, this kind of coherent search is not used 
when a DM value (or a narrow DM range) is not available.

Finally, in the case of two clusters (Liller~1 and Terzan~5, the
latter containing 2 already known MSPs), 
we have established a collaboration with Scott Ransom, whose
code \cite{rans01} looks promising for detecting MSPs orbiting
in ultra-close binaries, whose orbital period is shorter than the
typical duration of an observation.

\section*{Results to date}

So far we have discovered 12 new pulsars in 6 
globular clusters, none of which had previously known
pulsars associated with them. Seven of these pulsars are members of
binary systems, and 6 of them have relatively high DM values.
One pulsar follows a highly eccentric orbit and another one 
is eclipsed for a large fraction of the orbital period.
Their preliminary parameters are reported in Table~1.

\begin{center}
\vskip 1.0truecm
{\large{\bf Table 1: }}{\sc PARAMETERS OF THE 12 NEW PSRs IN GCs}
\vskip 0.5truecm
\begin{tabular}{c|llrlll}
\hline \\ 
CLUSTER &~PULSAR & ~Period & DM~~~~ & ~~$P_{b}$$~^{\dagger}$ 
&~a~${\rm sin}i$$~^{\ddagger}$ & ~M$_{c}^{min}$$~^{\diamond}$ \\

 &          & ~~(ms)      & (cm$^{-3}$pc) & (days) & (light-s) & (M$_\odot$) 
\\  
\\ 
\hline \\
                          & J1701$-$30A & ~~5.241    & 114~~~~      & 3.80   & 3.48   &   0.19  
\\ 
\\
{NGC 6266}& J1701$-$30B   & ~~3.593    &  114~~~~     &  0.14      & 0.25      &   
0.12  \\ 
\\
                          & J1701$-$30C & ~~3.806    & 114~~~~      & 0.21       & 0.19      &  0.07    \\ 
\\
\hline 
\\
{NGC 6397} & J1740$-$5340 & ~~3.650    & 72~~~~       & 1.35  & 1.66   & 0.19   \\ 
\\
\hline \\
{NGC 6441} & J1750$-$37   & 111.609    & 233~~~~      & 17.3 &  24.4   &  0.50 
\\ 
           &              &            &              & $e$=0.71 &     &      
\\
\hline \\ 
{NGC 6522} & J1803$-$30   & ~~7.101    & 192~~~~      & single &    &    \\  
\\
\hline \\
{NGC 6544} & J1807$-$2459 & ~~3.059    & 134~~~~      & 0.07  &0.012 & 0.009    \\ 
\\
\hline \\
                          & J1910$-$59A  & ~~3.266   & 34~~~~       & 0.86   &1.27    &  0.19   
\\ 
\\
                          & J1910$-$59B  & ~~8.357    & 34~~~~       & single &     &   \\ 
\\
{NGC 6752} & J1910$-$59C  & ~~5.277    & 34~~~~       & single &     &    \\ 
\\
                          & J1910$-$59D & ~~9.035    & 34~~~~       & single &      &     \\  
\\
                          & J1910$-$59E & ~~4.571    & 34~~~~       & single &      &     \\ 
\\  
\hline 
\end{tabular}
\end{center}
\vspace{0.3truecm}
{\small{$~~~~~~~~~~\dagger~=$ Orbital period\\
       $~~~~~~~~~~~\ddagger~=$ Projected semi-major axis of orbit\\
       $~~~~~~~~~~~\diamond~=$ Minimum companion mass, assuming that 
        the pulsar mass is $1.4~{\rm M_\odot}$}}

\subsection*{Five millisecond pulsars in NGC 6752}

NGC 6752 is classified as a core-collapsed cluster with evidence
of mass segregation \cite{fcbro97}. Rubenstein \& Bailyn \cite{rb97}
estimated a binary fraction for main sequence stars in the range 
15\%$-$38\% inside the inner core radius ($< 11''$), decreasing to less 
than 16\% beyond that. At least 6 dim X-ray sources have been identified
on the basis of ROSAT pointings \cite{vj00}. This is one of the clusters
with the more precise distance measurement: $4.1~{\rm kpc}\pm 5\%$,
obtained fitting the white dwarf (WD) sequence \cite{renzini96}.

In this cluster we first 
discovered a 3.26 ms binary pulsar, 
PSR~J1910$-$59A \cite{dlm+01}. This pulsar has
a relatively low DM = 34 cm$^{-3}$pc, and scintillates markedly, 
similar to the pulsars in 47~Tuc. The orbital solution gives $P_b = 20.6$ hr 
and a minimum mass for the companion of $M_{c}^{min}=0.19~{\rm M_\odot}$
(assuming, as everywhere in this paper, that the pulsar mass is 
$1.40~{\rm M_\odot}$). According to these parameters, it resembles a
MSP$+$He-WD binary similar to those discovered in 47~Tuc \cite{clf+00,egh+01}.

Amplification due to scintillation helped in the detection of four additional 
MSPs in the same cluster (Table~1). All of them are isolated 
with spin periods in the range $4.6-9.0$ ms.

\subsection*{Three binaries in NGC 6266}

Another GC in which we have detected more than one MSP is NGC 6266.
Its distance is estimated with the fitting of horizontal branch (HB) 
stars ($\sim 6.7$ kpc \cite{bbmp96}), but the high level of reddening
makes this measurement largely uncertain. It is classified as core-collapsed
but with ambiguities \cite{h96}. Re-examining all the ROSAT pointings,
Verbunt recently found a weak X-ray source, possibly 
interpreted as a quiescent soft X-ray transient\cite{v01}.

The first MSP discovered here was PSR~J1701$-$30A \cite{dlm+01}, with a spin 
period of 5.24 ms, an orbital period of 3.8 days, and a mass function 
indicating a minimum companion mass of 0.19 M$_{\odot}$. Apart
from the longer $P_b$ this binary is similar to PSR~1910$-$59A, whereas 
the two other new systems, PSRs~J1701$-$30B and J1701$-$30C, 
belong to the class of short-binaries with orbital periods of 3.8 hr and 
5.2 hr (Table~1). The close orbit of PSR~J1701$-$30B 
and the relatively high minimum companion mass cause significant variations
of the acceleration term $a$ during an observation. In this case, only the
use of code developed at Bologna for searching both $a$ and the derivative
of the acceleration, ${\rm d}a/{\rm d}t$, allowed the recovery of enough signal
from this weak source at the confirmation stage (see Figure~1).

\subsection*{An ultra-compact binary in NGC 6544}

NGC 6544 is one of the smallest GCs known: half of its mass is confined
within $1.3$ pc (at a distance of $\sim 2.6$ kpc, measured by HB-star-fitting
\cite{a83}). It displays the highest central luminosity density 
($\rho_0=5.75$ in logarithm of solar luminosities per cubic pc) among the
GCs having an associated pulsar (it ranks second 
among the entire GC population \cite{h96}). 

The pulsar discovered, PSR~J1807$-$2459 \cite{dlm+01,rans01}, has a
fairly high flux density and was not detected in
previous searches because of the relatively high DM = 134 cm$^{-3}$pc.
Timing performed at Jodrell Bank shows that it is part of an
extremely compact binary, with an orbital period of only 1.7 hours 
(the second shortest known) and a projected semi-major axis of only
$0.56~{\rm R_\oplus}$. The corresponding minimum companion mass is only 
0.009 M$_{\odot}$ or about 10 Jupiter masses (the second smallest known) 
and the orbital separation is 
$(88/\sin i)~{\rm R_\oplus}=(0.8/\sin i)~{\rm R_\odot}.$ If the 
orbital inclination angle $i$ is small, this system could be similar to the
close systems (MSP+WD or MSP+light not degenerate companion)
seen in 47~Tuc \cite{clf+00}, while if $i\sim 60^\circ$ (the a priori
median of the possible inclinations) the companion could be a sub-stellar
object: a brown dwarf or a planet.

\begin{figure}[ht]
\centerline{\epsfig{file=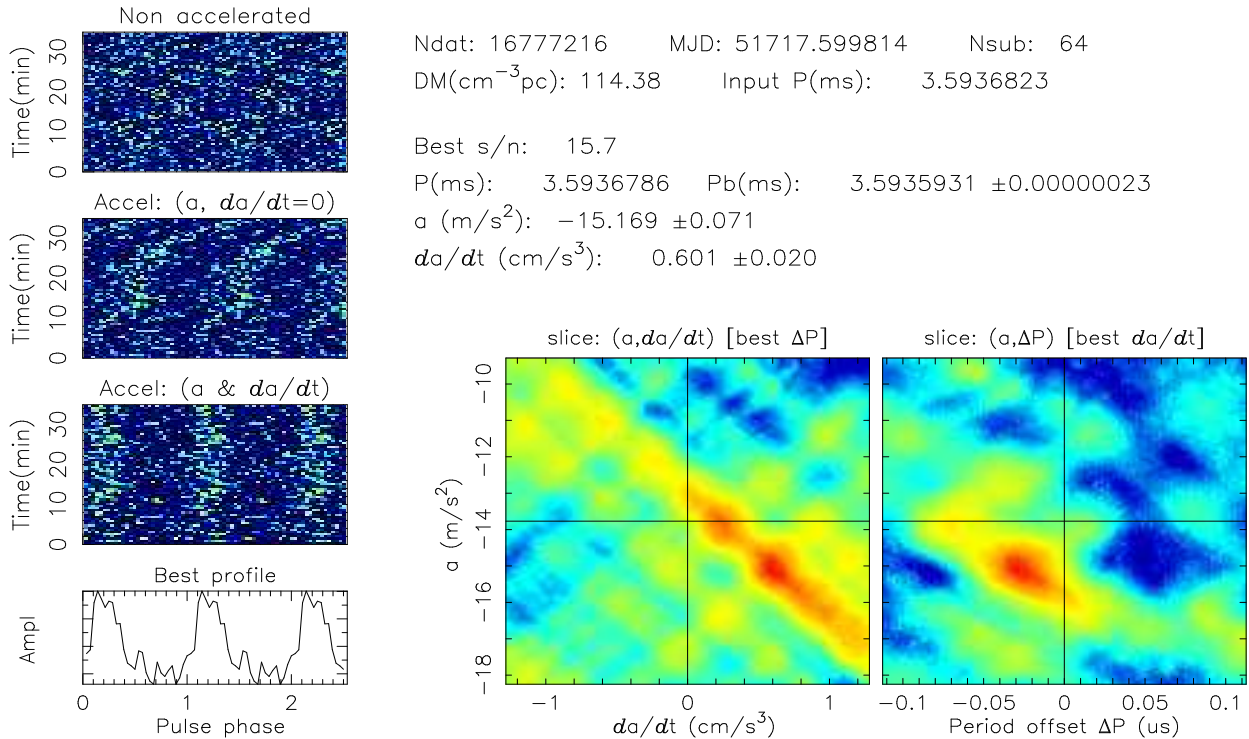,height=4.5in,width=6.7in}}
\vskip 0.3truecm
\caption{Confirming plot for PSR~J1701$-$30B. In the upper left color
panel is displayed the time-resolved signal from the source folding 
the 30-min data set at the nominal candidate spin period $P$ and
DM. The middle left panel shows the same signal after 
including a constant acceleration term $a$ during the integration. 
The lower left color plot presents the time-resolved signal folded with
the best values of $a$ and ${\rm d}a/{\rm d}t$.
After these corrections, the pulse profile is easily recognizable (panel
labeled with {\it Best profile\/}) with a good s/n=15.7. The two square panels
in the lower section of the figure contain color-scale confidence plots
for the three parameters searched: $a$, ${\rm d}a/{\rm d}t$ and the period
offset $\Delta P$ with respect to the original detection.}
\label{figA}
\end{figure}

\subsection*{A peculiar eclipsing pulsar in NGC 6397}

NGC 6397 is one of the most promising targets for searching MSPs in
GCs: in fact, it is one of the closest clusters, at a distance of 
$2.6~{\rm kpc}~\pm 6\%$ \cite{rg98}, and probably has a collapsed core
with hints of mass segregation \cite{ksc95}. It lies in fourth place
in the list of GCs ranked according to central luminosity \cite{h96}.
Moreover, it contains $\sim 20$ X-ray sources detected with {\sl Chandra}
within $2\arcmin$ of the cluster center, 8 of which
probably identified with CVs \cite{ghemc01}.  
 
\begin{figure}[ht]
\centerline{\epsfig{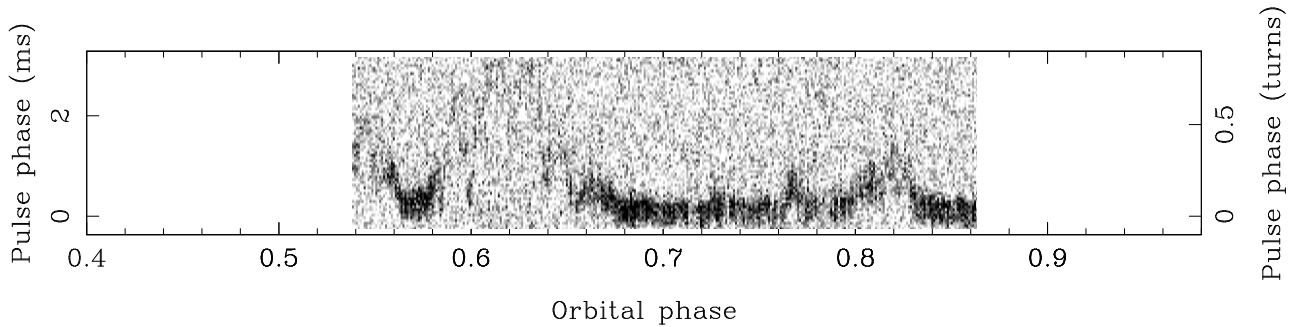}}
\vskip 0.3truecm
\caption{Greyscale plot of the signal intensity at 1.4~GHz as function of 
orbital phase (between phases 0.54 and 0.86) and pulsar phase 
for a 11 hr long observation of PSR J1740$-$5340 taken far from
the nominal eclipse region (whose ingress is at orbital phase $\sim 0.05$
and egress at orbital phase $\sim 0.45$). The data are folded at
constant period in sub-integrations lasting 120 s each. Darker regions
correspond to higher signal intensity and thus roughly locate the 
average phase of the pulse in each sub-integration. The reference phase
for the arrival of the pulses is set to 0. Upward shifts of the darker 
regions are a signature of delays occurring in the time of arrival of the 
pulses. The observation started at 15:02 UT on 2001 March 12 [34].}
\label{figB}
\end{figure}  

In this cluster we have discovered PSR~J1740$-$5340, a binary MSP with 
a spin period of 3.65 ms \cite{dlm+01}. It displays eclipses at 
1.4 GHz for more than 40\% of the orbital phase. This phenomenon 
is not uncommon (10 eclipsing systems containing a MSP are known
\cite{clf+00,fbb+90,l90,sbl+96,nat00,dpmsl01}) 
and in the case of PSR~B1744$-$24A\cite{nt92} the eclipses 
show duration and irregularities similar to those of PSR~J1740$-$5340
However this new system is $2-3$ times wider (with an orbital separation
of $\sim 6.5~{\rm R_\odot}$) than any other known eclipsing pulsar binary,  
and has a heavier minimum mass for the companion ($>$ 0.19 M$_{\odot}$) than
any known eclipsing system. In addition, the radio signal
exhibits striking irregularities (delays and intensity variations)
over a wide range of orbital phases (see Figure~2), indicating
that the MSP is orbiting within a large envelope of matter released
from the companion with a high mass loss rate.
These characteristics challenge the evaporation models 
from a degenerate companion, suggesting more likely that the companion 
is a bloated main-sequence star or the remnant of the star 
that spun up the MSP\cite{dpmsl01}.
No millisecond pulsar with such peculiar companion is known to date.
The detection of the pulsar in the X-ray band\cite{ghemc01} and the
probable optical identification of the companion\cite{fpds01} open also
the possibility of extended follow-up observations 
which will probe the eclipse mechanism and the true nature of the companion 
in this very interesting system.

\subsection*{An isolated millisecond pulsar in NGC 6522}

At a distance of $7.8~{\rm kpc}$ \cite{tpgrs98} and with Galactic coordinates
$l=1.02^\circ$ and $b=-3.93^\circ$, NGC 6522 is one of the GCs located near 
the bulge of the Galaxy, only $600~{\rm pc}$ away from the center. 
It is classified as core-collapsed \cite{h96}, but the most 
interesting dynamical feature resides in its orbit, which recently 
($\sim 2\times 10^6$ yr ago) experienced a close encounter 
(at $\sim 400-500$ pc) with the Milky Way center. Hence, NGC 6522 could
have experienced dynamical shocks due to the time-variable gravitational
potential of both the bulge and the inner disk of the Galaxy \cite{go97}. 

So far in this survey, this is the only GC for which 
the only pulsar detected has been an isolated MSP, with a spin period
of 7.1 ms\cite{pdm+01}. PSR~J1803$-$30 is in fact relatively 
strong and the previous surveys probably missed it due to its 
high DM = 192 cm$^{-3}$pc.  

\subsection*{\vspace{-0.05truecm}A highly eccentric binary pulsar in NGC 6441}

The most recent discovery occurred in the cluster NGC 6441. Its
large distance (11.2 kpc) has been determined using RR Lyrae as
calibrators \cite{lrww99}. It has a high central luminosity density
but it is not classified as core-collapsed \cite{h96}. It hosts a 
bright low mass X-ray binary, XB~1746$-$371, whose optical companion
has been likely identified at $6''$ from the cluster center \cite{damd98}.
It also contains one of the rare GC planetary nebulae, JaFu 2 
\cite{jf97}.

Here we have found a binary pulsar, PSR~J1750$-$37,
which has the second largest DM (233 cm$^{-3}$pc)
and the longest spin period, 111.6 ms, among all the known 
GC pulsars (PSR~B1718$-$19 has an even longer period,
but its inclusion in NGC 6342 is unlikely\cite{ka96}).
After some months of timing, we have obtained an
orbital solution (see Table~1) indicating that the   
system is large ($P_b=17.3$ days), and in a highly eccentric orbit
with $e = 0.71$\cite{pdm+01}.
The minimum companion mass ($M_{c}^{min} > 0.5$) suggests it
could be a binary comprising a massive WD or a second NS.

\section*{Conclusions}

After two years of observations and processing of data,
we have discovered 12 new pulsars in 6 GCs, increasing the total
sample of such pulsars by about 25\%: there are now 54. We are
timing all the new sources and within a few months will derive
precise celestial coordinates for most of them. 
With these discoveries, we have also increased by about 50\% the number 
of clusters (now 18) having at least one associated pulsar,
and for which we have therefore also a good estimate of the DM. 
The consequent reduction 
in the number of unknown search parameters will make computationally 
feasible a future even deeper investigation of the pulsar content of 
these stellar systems, allowing a more reliable study of its dependence 
on other GC parameters such as mass, concentration, central stellar 
density, dynamic state and orbit in the Galaxy. 

To date we have not obtained any pulsar detection in about 30 observed GCs.
Our plan is to re-observe all of them. Indeed, the present experiment 
suggests that, besides sensitivity and powerful search algorithms, 
a key strategy in a search for GC-MSPs is to devote a large amount 
of observing time to each target. Scintillation in low DM 
clusters, long eclipses, and unfavorable orbital phases 
in the case of ultra-short binaries, might easily prevent detection 
during a single observation. For a subset of promising GCs we
will also start a massive multidimensional coherent search in 
acceleration for a very large number of DM trials, exploiting
the computing facilities available at the Cineca Supercomputing Center
near Bologna.

Finally we will collect data for the $\sim 20$ GCs
for which we have not yet performed observations: among them
are the GCs in our list containing already known MSPs.

\end{document}